\documentstyle[12pt]{article}
\begin{document}
\begin{center}
{\bf{E.C.G. SUDARSHAN Memorial Lecture{\footnote{Dr.E.C.G.Sudarshan Memorial Lecture on 13 May 2019, at 
Town Hall, Ernakulam}}}}
\end{center}

\vspace{0.5cm}

\begin{center}
R.Parthasarathy{\footnote{e-mail: sarathy@cmi.ac.in}} \\
Chennai Mathematical Institute \\
Siruseri, Chennai 603 103 \\
\end{center}

\vspace{0.5cm}

{\noindent{\it{Abstract}}}

\vspace{0.5cm}

In this invited talk, I discuss four important contributions of E.C.G. Sudarshan, among many.
They are the V-A theory of weak interaction, Sudarshan - Glauber representation, tachyons and 
the quantum zeno effect. 

\vspace{1.0cm}

Honorable Vice Chancellor of Cochin University of Science and Technology, Prof. K.N.
Madhusoodanan, respected President of Swadeshi Science Movement - Kerala, Dr. V.N.
Sanjeevan, respected Former Director of CMFRI, Dr. N.G.K. Pillai, respected Secretary 
of Swadeshi Science Movement - Kerala, Dr. A.R.S. Menon and respected Secretary of 
Vijnana Bharati, Shri. P.A. Vivekananda Pai, I am honoured to  give Dr.E.C.G.
Sudarshan Memorial Lecture. I am thankful to the Swadeshi Science Movement - Kerala 
for inviting me here and to Shri. Vivekananda Pai for the excellent hospitality. 
Usually Memorial Lectures will be on some special topics of the common interest.
I chose to talk on Dr.E.C.G. Sudarshan and his seminal works as these will be of 
common interest and homage to the great theoretical physicist. 

\vspace{0.5cm}

Born on 16 September 1931 in Pallam, Kerala, Ennackal Chandy George Sudarshan became 
one of the top few Theoretical Physicists in the world. It was realized around 1960's 
that only a handful of physicists could unravel the mysteries of Particle Physics and 
Sudarshan was one among them.

\vspace{1.0cm}

After his early education at CMS College, Kottayam, he joined Madras Christian College 
(1951) and received Master's degree from the University of Madras in 1952. He joined 
TIFR, Mumbai and after a brief period, he went to the University of Rochester, USA, for 
his Ph.D under the guidance of Prof.R.E. Marshak. After completing Ph.D he went to 
Harvard University as a Post-doc of J.Schwinger, one of the three to receive Nobel Prize 
for Quantum Electro Dynamics. In 1969, he moved from Rochester to the University of Texas,
Austin and remained there till his end which came on 13 May 2018.

\vspace{1.0cm}

When in Rochester, he frequently visited India. In particular he spent considerable time 
in MATSCIENCE (IMSc) giving lectures on Particle Physics. In 1970, he was Sir.C.V.Raman
visiting Professor to University of Madras giving lectures on Quantum Field Theory and 
that is when I had the opportunity to meet him. 

\vspace{1.0cm}

Then, he was appointed as the Director of the Institute of Mathematical Sciences, Madras,
invited by the Prime Minister Srimathi Indira Gandhi. He was the Director of IMSc from 
1984-89. I was an Associate Professor at IMSc then and interacted with him closely on 
research topics, particularly Kaluza-Klein theory of higher dimensional gravity. 

\vspace{1.0cm}

He made very important and trend setting contributions in Theoretical Physics. Among his 
contributions, I will focus on \\

\noindent{$\bullet$} Originator of V-A theory of Weak Interactions \\
\noindent{$\bullet$} Sudarshan-Glauber representation of coherent light \\
\noindent{$\bullet$} Tachyons \\
\noindent{$\bullet$} Quantum Zeno Effect \\  
the main reason being "each one deserves the award of Nobel Prize to Sudarshan."

\vspace{1.0cm} 

\noindent{\bf{V-A Theory of Weak Interactions (1957)}}

\vspace{0.5cm}

In nature, there are Four Fundamental Interactions. They are:\\

\noindent (1) Gravitational Interaction - responsible for planetary motions and Cosmology.\\

\noindent (2) Electromagnetic Interaction - responsible for atom stability, Solids and Chemistry.\\

\noindent (3) Strong Interaction - responsible for holding neutrons and protons inside atomic nucleus.\\

\noindent (4) Weak Interaction - responsible for the decay of unstable particles like neutron.  \\

\vspace{0.5cm}

Newton gave the rules for Gravitational Interaction and perfected by Einstein.\\

Maxwell gave the rules for electromagnetic interaction and the quantum theory was successfully 
carried out by Dirac, Feynman, Schwinger and Tomonaga. \\

Strong interaction was initially explained by Yukawa. The Quark Model of Gell-Mann explained the 
static properties of strongly interacting particles and the quantum field theory of strong 
interactions later developed as Quantum Chromo Dynamics - color force. \\

The situation in Weak Interactions was involved!  

\vspace{1.0cm}

First, in the beta decay of heavy nucleus, a manifestation of Weak Interaction, the energy spectrum 
of $\beta$-particles (electrons) was {\it{continuous}}. One expected a discrete spectrum due to 
quantized energy levels of the initial and final nuclei. Great minds like Bohr suspected energy 
conservation! Pauli in 1930 suggested that along with electrons, massless, neutral and spin half 
particles could be emitted - Neutrino. Such particles were discovered. 

\vspace{1.0cm}

Second, to explain $\tau - \theta$ puzzle, Lee and Yang in 1956 [1] suggested boldly that 
Parity (spsce reflection symmetry) should be violated in Weak Interactions. Next year, 
C.S.Wu and her co-workers discovered experimentally parity violation in the beta decay of 
$^{60}Co$ nucleus [2]. So Parity violation in Weak Interactions was established. 

\vspace{1.0cm}

What is the rule or form of the Hamiltonian for Weak Interactions? Not Clear!! In 1933, Fermi [3]  
proposed a vector theory of beta decay in analogy with electromagnetic interaction. 
\begin{eqnarray}
{\cal{L}}_{Fermi}&=&-\frac{G_F}{\sqrt{2}} {\bar{\psi}}_p{\gamma}^{\mu}{\psi}_n\ {\bar{\psi}}_e
{\gamma}_{\mu}{\psi}_{\nu}+h.c. \nonumber 
\end{eqnarray}
The weak current is a vector (Lorentz) and so conserves parity. This form explained 
$\bigtriangleup J=0$ (no spin flip) process. The dominant term of the current 
$j_{\mu}^{(n\rightarrow p)}={\bar{\psi}}_p{\gamma}_{\mu}{\psi}_n$ causes no spin-flip 
transition - Fermi Transition.
 
But, there are spin flip transitions observed. Suppose we consider $J_{\mu}^{(n\rightarrow p)}=
{\bar{\psi}}_p{\gamma}_{\mu}{\gamma}_5{\psi}_n$, then the dominant term will be 
${\phi}_p^*\vec{\sigma}{\phi}_n$, causing spin flip transition - Gamow Teller transition like 
$^{12}B(1^+)\rightarrow ^{12}C(0^+)+e^-+{\tilde{\nu}}_e$. So, the axial vector form is needed 
to explain spin flip transition. Thus to account for Fermi and Gamow-Teller transitions, one 
can try a linear combination of ${\bar{\psi}}_p{\gamma}_{\mu}{\psi}_n$ and ${\bar{\psi}}_p
{\gamma}_{\mu}{\gamma}_5{\psi}_n$; V and A mixture. But is it V-A or V+A?  

\vspace{1.0cm}

Here the neutrino helps. The massless Dirac equation suggests 
\begin{eqnarray}
{\psi}_{{\nu}_L}&=&\frac{1}{2}(1-{\gamma}_5){\psi}^{(2)}. \nonumber 
\end{eqnarray}
Experimentally, neutrino was found to be left handed. So suggestive of V-A, ${\bar{\psi}}_p
{\gamma}_{\mu}(1-{\gamma}_5){\psi}_n$. However, there was utter confusion prevailed before 1956.
Physicists tried all possible linear combinations, like vector, axial vector, scalar, tensor etc.
for ${\bar{\psi}}_p\cdots {\psi}_n$.

\vspace{1.0cm}

Sudarshan in 1956 boldly suggested V-A theory as 
\begin{eqnarray}
{\cal{L}}&=&g{\bar{\psi}}_1{\gamma}_{\mu}(1-{\gamma}_5){\psi}_2\ {\bar{\psi}}_3{\gamma}^{\mu}
(1-{\gamma}_5){\psi}_4, \nonumber 
\end{eqnarray}
a V-A form for the weak current and with maximal parity violation. 

\vspace{1.0cm}

When he discussed with Marshak, they decided to propose this in a Conference in 1956. As Marshak was 
presenting his work, and Sudarshan could not present as he was a graduate student, Marshak requested 
P.T.Mathews to present their work and this did not happen! "When I was sitting in the Conference with 
the correct theory, people tried all possible explanations", he told me in 1986. Next year, in 
Padua meeting  Sudarshan-Marshak presented their epoch making result - {\it{the discovery of 
the law of one of the four fundamental forces of nature}} [4].

\vspace{0.5cm}

Sudarshan was aware that among the experimental data,four were disagreeing with his theory. They were 
$e-\nu$ angular correlation in $^6He$ decay, signs of electron polarization in muon decay, frequency 
of electron mode in pion decay and asymmetry in polarized neutron decay. However, he was convinced of 
the correctness of his theory and insisted that these experiments should be repeated. All the four 
experiments were repeated and the results changed, falling in line with his V-A predictions. \\

{\it{Thus the V-A theory of weak interaction was firmly established.}}

\vspace{1.0cm}

Feynman and Gell-Mann [5] took a different route and came up with the V-A theory a year later, Phys.Rev.
109 (1958) 193. 

\vspace{1.0cm}

Is that all? The V-A form for the weak current plays a {\underline{fundamental role}} in constructing the 
gauge theory of weak and electromagnetic interactions - unified theory. It is known that from the 
free Dirac Lagrangian density
\begin{eqnarray}
{\cal{L}}_D&=&\tilde{\psi}i{\gamma}^{\mu}{\partial}_{\mu}\psi, \nonumber 
\end{eqnarray}
imposing local gauge invariance $\psi\rightarrow e^{i\theta(x)}\psi$, one obtains the e.m interaction 
$j^{\mu}A_{\mu}$ with $j^{\mu}=\tilde{\psi}{\gamma}^{\mu}\psi$ - Vector current. Now defining 
${\psi}_L=\frac{1}{2}(1-{\gamma}_5)\psi$, one starts from 
\begin{eqnarray}
{\cal{L}}&=&{\tilde{\psi}}_Li{\gamma}^{\mu}{\partial}_{\mu}{\psi}_L, \nonumber 
\end{eqnarray}
one finds after local gauge invariance for ${\psi}_L$, 
\begin{eqnarray}
j^{\mu}&=&{\tilde{\psi}}_L{\gamma}^{\mu}{\psi}_L=\frac{1}{2}\tilde{\psi}{\gamma}^{\mu}(
1-{\gamma}_5)\psi, \nonumber 
\end{eqnarray}
the V-A current amenable to gauge theory construction. Of course one needs to go beyond U(1) gauge 
theory. The unification of weak and e.m interaction is achieved by gauge group $SU(2)_L\times 
U(1)_Y$ in which 
\begin{eqnarray}
{\psi}_L&=&\left(\begin{array}{c}
{\nu}_e \\
e \\
\end{array}\right)_L
\ ;\  SU(2)_L\ \ doublet, \nonumber \\
{\psi}_R&=&e_R ;\ \ SU(2)_L singlet. \nonumber 
\end{eqnarray}
Such unification has been realized by S.Weinberg (1967) [6], A.Salam (1969) [7] and S.L.Glashow, J.Iliopoulos , 
L.Maiani (1970) [8]. 

\vspace{1.0cm}

Weinberg, Salam and Glashow were awarded Nobel Prize in 1979. First, the V-A form for the weak current, the 
theory of Sudarshan and Marshak, is so fundamental that it successfully works for the quark sector,
although the original theory involved nucleons. Second, the V-A form forms the {\it{Foundation of the 
electroweak theory}}, the edifice being raised by Weinberg and Salam. It is very painfully 
disappointing that the persons who laid the foundation were not awarded. 

\vspace{1.0cm}

It is worth recalling: Feynman to Marshak: "I hope some day we can get this straightened out and give 
Sudarshan the credit for priority he justly deserves"

\vspace{0.5cm}

S.Weinberg: ".... it took tremendous courage for Marshak and Sudarshan to propose.......that in 
fact the weak interaction was vector and axial vector.......Marshak and Sudarshan were the first 
to propose in 1957". 

\vspace{0.5cm}

Though the coveted award remained elusive (Sudarshan was nominated nine times), the fact the rule of 
one of the fundamental forces of nature was first discovered by Sudarshan and Marshak will be etched 
in  Theoretical Physics permanently. 

\vspace{1.0cm}

\noindent{\bf{Diagonal Representation in Quantum Optics (1963)}}

\vspace{0.5cm}

To describe quantum states close to classical beams of light, Sudarshan [9] in 1963 proposed the 
equivalence of semiclassical and quantum mechanical description of statistical light beams 
in Phys. Rev. Lett. (1963) - first one in the field of optics. Briefly, on account of the 
over completeness of coherent states (introduced by Schr\"{o}dinger) $|z\rangle$ , any density 
matrix $\rho$ can be expressed in the diagonal representation 
\begin{eqnarray}
\rho &=&\frac{1}{\pi}\int d^2z\ \phi(z)\ |z\rangle \langle z|. \nonumber 
\end{eqnarray}
A formal (singular though) expression for $\phi(z), z=re^{i\theta}$ is 
\begin{eqnarray}
\phi(z)=\sum_{n,n'=0}^{\infty}\frac{\sqrt{n!}\sqrt{n'!}}{(n+n')!}\langle n|\rho|n'\rangle \nonumber \\
\frac{1}{2\pi r}e^{r^2+i(n'-n)\theta}\left(-\frac{\partial}{\partial r}\right)^{n+n'}\delta(r), \nonumber 
\end{eqnarray}  
The quantum mechanical correlation function 
\begin{eqnarray}
G^{(1,1)}\rightarrow Tr(\rho a^{\dagger}a)=\frac{1}{\pi}\int d^2z \phi(z)z^*z, \nonumber 
\end{eqnarray}
$\phi(z)$ is known as Sudarshan-Glauber phase space function. Glauber published in the same year [10]. 

\vspace{1.0cm}

The optical equivalence theorem forms the central tool in quantum optics. Sudarshan established that 
$\phi(z)$ is universal for all states of e.m field. In fact, a q-generalization of Sudarshan's result 
has been worked out by myself and R.Sridhar [11] using q-boson coherent states.

\vspace{0.5cm}

Glauber shared 2005 Nobel prize with experimentalists J.L.Hall and T.W.Hansch. It is very unfortunate 
that Sudarshan who equally deserved the award was not given - a great injustice. This is the second time 
the award was unjustly denied. 

\vspace{1.0cm}

\noindent{\bf{Tachyons (1962)}}

\vspace{0.5cm}

A {\it{hypothetical particle}} traveling with speed exceeding the speed of light was proposed by 
Bilaniuk, Deshpande and Sudarshan in 1962 [12]. Independently, Feinberg [13] proposed such particles in 1967.
A.Sommerfeld, Ehrilch proposed earlier in 1904. As in the usual theory,
\begin{eqnarray}
E^2=(pc)^2+m^2c^4 &;& E=\frac{mc^2}{\sqrt{1-\frac{v^2}{c^2}}}. \nonumber 
\end{eqnarray}
When $v>c$, as $E$ must be real, $m$ becomes imaginary. or $m^2$ is negative. Such particles have 
not been observed till now. However, the concept prevails in Quantum Field Theory such as 
spontaneous symmetry breaking. In string theory, one encounters tachyonic modes and to eliminate them 
the space dimension is increased to $D=26$ or $D=10$. This trend setting proposal received 
considerable interest although not realized in nature. 

\vspace{1.0cm}

\noindent{\bf{Quantum Zeno Paradox (1977)}}

\vspace{0.5cm}

In 1977, Misra and Sudarshan [14] showed that if one checks sufficiently frequently whether the 
initial state $|\psi(t_0)\rangle$ has not decayed, then as the frequency of observation increased 
indefinitely, the state will not decay at all! In the case of an unstable particle, frequent 
measurement inhibits the decay. Experiments by Itano et.al [15] have confirmed the existence of 
Quantum Zeno Effect.

\vspace{0.5cm}

First, to give a feeling, consider an isolated physical system in a state ${\psi}_0$ at some 
initial time $t=0$. Let it be allowed to run for a very short time $t$ and is then observed. 
What is the probability that the physical system will still be in its initial state? Let $H$ 
be independent of time. Then at time $t$
\begin{eqnarray}
\psi(t)&=&e^{-iH\frac{t}{\hbar}}\ {\psi}_0. \nonumber 
\end{eqnarray}
Then the probability that it is in the initial state is 
\begin{eqnarray}
P&=&|\langle {\psi}_0|e^{-i\frac{Ht}{\hbar}}|{\psi}_0\rangle|^2. \nonumber 
\end{eqnarray}
If the time $t$ is very short , then 
\begin{eqnarray}
P&\simeq &|\langle{\psi}_0|1-i\frac{Ht}{\hbar}-\frac{1}{2}\frac{H^2t^2}{{\hbar}^2}+\cdots |{\psi}_0
\rangle|^2. \nonumber 
\end{eqnarray}
Keeping terms up to $t^2$, 
\begin{eqnarray}
P&\simeq &\{ 1-i\langle H\rangle \frac{t}{\hbar}-\frac{1}{2}\langle H^2\rangle \frac{t^2}{{\hbar}^2}\} \nonumber \\
      & & \{1+i\langle H\rangle \frac{t}{\hbar}-\frac{1}{2}\langle H^2\rangle \frac{t^2}{{\hbar}^2}\}, \nonumber \\
        &\simeq &1-\langle H^2\rangle \frac{t^2}{{\hbar}^2}+{\langle H\rangle}^2\frac{t^2}{{\hbar}^2}, \nonumber \\
       &\simeq &1-(\bigtriangleup H)^2\frac{t^2}{{\hbar}^2}, \nonumber 
\end{eqnarray}
where $(\bigtriangleup H)^2=\langle H^2\rangle-{\langle H\rangle}^2$. 

\vspace{1.0cm}

Thus, the change begins slowly. Suppose the system is observed at $N$ equally spaced intervals over a time $T$. 
Each measurement perturbs the system a little and so successive values of $\bigtriangleup H$ will not all be 
the same. But since the measurements are independent events, probabilities multiply  and with $t=\frac{T}{N}$,
the probability that the system remains in ${\psi}_0$ is 
\begin{eqnarray}
\bar{P}&=&\prod_{i=1}^N \{1-\left(\bigtriangleup H_i\right)^2\frac{T^2}{{\hbar}^2N^2}\}. \nonumber 
\end{eqnarray}
Taking logarithm 
\begin{eqnarray}
\ln \bar{P}&=&\sum_{i=1}^N\ln \{1-\left(\bigtriangleup H_i\right)^2\frac{T^2}{{\hbar}^2N^2}\}, \nonumber \\
           &=&\sum_{i=1}^N-\left(\bigtriangleup H_i\right)^2\frac{T^2}{{\hbar}^2N^2}+\cdots . \nonumber 
\end{eqnarray}
Then 
\begin{eqnarray}
\bar{P}&=&e^{-\bar{(\bigtriangleup H)^2}\frac{T^2}{{\hbar}^2N^2}}, \nonumber 
\end{eqnarray}
where $\bar{(\bigtriangleup H)^2}$ is the average. In the limit of large $N$, that is, if the system is observed 
continuously, $\bar{P}=1$, and the system never changes!! 

\vspace{1.0cm}

{\it{Colorado group in 2009}} [15] realized the occurrence of Quantum Zeno Effect in induced transitions 
between two quantum states. Itano et.al., tested the inhibition of the induced radio frequency transition 
between two levels of Be ion. Suppose the ion is in level 1 at $t-0$. A RF field of frequency 
$\Omega =\frac{E_2-E_1}{\hbar}$ is applied. This creates a state which is a coherent superposition of 
states 1 and 2. An on-resonance pulse of duration $T=\frac{\pi}{\Omega}$ takes 1 to 2. If $P_2(t)$ is 
the probability at time $t$ for the ion to be in level 2, then $P_2(t)=1$. In the experiment,
N measurement pulses are applied which connect 1 to 3 through an optical pulse each time 
with in T that is ${\tau}_k=\frac{kT}{N}$, k=1,2,$\cdots$ N. At the end of N measurements, that is 
at the end of RF pulse at time t,
\begin{eqnarray}
P_2(t)&=&\frac{1}{2}(1-{\cos}^N(\frac{\pi}{N})). \nonumber 
\end{eqnarray}
For large $N$, that is in the limit continuous measurement (of 1), $P_2(t)\sim 0$. The system is 
frozen in level 1. This confirms Quantum Zeno Effect. 

\vspace{1.0cm}

I have sketched four important contributions of Sudarshan, trend setting ones. In particular, the 
V-A theory, diagonal representation of density matrix and QZE, each one richly deserves the Nobel 
Prize. Though he was nominated nine times, it is puzzling why it did not materialize. 

\vspace{1.0cm}

Sudarshan was very considerate person. I found that he never hesitated to listen to my physics 
research and offered crucial points to better my approach. He became very much involved in the 
Vedic Philosophy. He had detailed discussions with J.Krishnamoorthy. When I observed him 
during 1984-89, he was like a Karma Yogi, deriving immense pleasure and satisfaction by his 
work, with out worrying about the rewards. I wish to end up with a human aspect of him. In November 
1985, heavy rains in Madras affected my house and my family had to move out. We went to IMSc 
guest house for shelter. Prof. Sudarshan generously offered his quarters to share and hosted us 
for a week, a gesture that I can never forget. 

\vspace{0.5cm}

{\noindent{\bf{References}}}

\vspace{0.5cm}

\begin{enumerate}
\item T,D, Lee and C.N. Yang, Phys.Rev. {\bf{104}} (1956) 254.
\item C.S. Wu, E. Ambler, R.W. Hayward, D.D. Hoppes and R.P. Hudson, Phys.Rev. {\bf{105}} (1957) 1413.
\item E. Fermi, Il Nuovo Cimento. {\bf{9}} (1934) 1.
\item E.C.G. Sudarshan and R.E. Marshak, {\it{The Nature of Four - Fermion Interaction}}, Proc. of the 
      conference on Mesons and Newly discovered Particles, Padua - Venice, Sept.1957, reprinted in 
      "Developments of the theory of weak interactions", P.K.Kabir (Ed), Gordon and Breach, NY, 1964.
\item R.P. Feynman and M. Gell-Mann, Phys.Rev. {\bf{109}} (1958) 193.
\item S. Weinberg, Phys.Rev.Lett. {\bf{19}} (1967) 1264.
\item A. Salam, Proc. of 8-th Nobel Symposium. Ed. N.Svartholm (Almquist and Wiksell, Stockholm) 1968, p.367.
\item S.L. Glashow, J. Iliopoules and L. Maini, Phys.Rev. {\bf{D2}} (1970) 1285.
\item E.C.G. Sudarshan, Phys.Rev.Lett. {\bf{10}} (1963) 277.
\item R. Glauber, Phys.Rev. {\bf{131}} (1963) 2766.
\item R. Parthasarathy and R. Sridhar, Phys.Lett. {\bf{A305}} (2002) 105. 
\item O.M.P. Bilaniuk, V.K. Deshpande and E.C.G. Sudarshan, Am.J.Phys. {\bf{30}} (1962) 718.
\item G. Feinberg, Phys.Rev. {\bf{159}} (1967) 1089.
\item B. Misra and E.C.G. Sudarshan, J.Math.Phys. {\bf{18}} (1977) 756.
\item W.M. Itano, D.J. Heinzen, J.J. Bollinger and D.J. Wineland, Phys.Rev. {\bf{A41}} (1990) 2295.
\end{enumerate}

\end{document}